
\def\v{\vskip 2mm}
\def\vv{\v\v}
\def\vvv{\v\v\v}

\def\noi{\noindent}
\def\no{\noindent}

\def\section#1{\vskip 8mm \noi {\bf #1} \vskip 8mm}
\def\subsection#1{\vskip 6mm \noi {\bf #1} \vskip 6mm}

\def\sub#1{\vskip 6mm \noi {\bf #1} \vskip 4mm}

\def\r{\hangindent=1pc  \noindent}

\def\ce{\centerline}

\def\endpage{\vfil\break}

\def\kms{km s$^{-1}$}

\def\deg{$^\circ$}

\def\vlsr{$V_{\rm LSR}$}

\def\Vlsr{V_{\rm LSR}}

\def\kms{km s$^{-1}$}
\def\Msun{M_{\odot \hskip-5.2pt \bullet}}

\def\Deg{^\circ}
\def\deg{$^\circ$}

\def\htwo{H$_2$}

\def\ta{$T^*_{\rm A}$}

\def\lv{$(l, V)$}
\def\bv{$(b, V)$}

\ce{\bf Galactic Center Molecular  Arms, Ring, and Expanding Shells. II}
\ce{\bf Expanding Molecular Shell}

\vvv
\ce{Yoshiaki SOFUE}

\vv
\ce{Institute of Astronomy, The University of Tokyo, Mitaka, Tokyo 181, Japan}
\ce{ sofue@mtk.ioa.s.u-tokyo.ac.jp}

\vv
\ce{({\it to appear in PASJ, 1995})}
\vvv
\ce{\bf Abstract}  \vv

We investigate the three-dimensional structure of the so-called
``200-pc expanding molecular ring" in the Galactic Center by
analyzing $(b, \Vlsr)$ diagrams of the $^{13}{\rm CO}(J=1-0)$ line
emission from  the Bell-Telephone Laboratory survey.
We show that the \bv\ features can be fitted by a spheroidal shell
pinched at the equator (dumbbell-shaped shell),
which we call the expanding molecular shell (EMS).
The radius is about 180 pc and the vertical extent is more than $\pm50$ pc.
The shell is expanding at 160 \kms, and is rotating at 70 \kms.
The association of the radio continuum emission indicates  that the
EMS is a mixture of molecular and ionized hydrogen gases.
An extended hot (X-ray) plasma appears to fill the interior of the shell.
The estimated molecular mass  of the shell is $\sim 10^7\Msun$
and its kinetic energy $\sim 2\times 10^{54}$ ergs.

\vv\no{\bf Key words}:  Expanding gas -- Galactic center -- Galaxy --
Interstellar matter --  Molecular gas.

\vv
\section{1. Introduction}

The molecular gas in the Galactic Center region
is generally confined in a thin dense layer of a few tens
of pc thickness composing the nuclear molecular disk,
while a small amount of out-of-plane gas is present
extending up to $\sim 50$ pc above the disk
(see Paper I and the literature cited therein).
In paper I (Sofue 1995) we have reanalyzed the longitude-velocity \lv
diagrams of the $^{13}$CO line data of Bally et al (1987, 1988),
and have shown the presence molecular arms and ring structure within the
nuclear disk.

Besides the disk components, the central 200 pc region is characterized
by the  ``200-pc expanding  molecular ring'', which exhibits high
positive- and negative-velocity features consisting of an `ellipse' in the
\lv\ diagram (Scoville 1972: Kaifu et al 1972, 1974).
It has been suggested  that this structure is
vertically extended for about 100 pc perpendicularly to the galactic plane,
and its cylindrical structure has been pointed out (Sofue 1989).
Alternatively, it might be a part of a ``parallelogram'' in the
\lv\ plot (Binney et al 1991), or
this feature might be an inner part
of an expanding tilted disk of 2 kpc radius (Liszt and Burton 1978).

In this paper, we  revisit this prominent feature
and investigate its vertical structure based on \bv\ diagrams
of the $^{13}{\rm CO}~(J=1-0)$ and CS lines observed by Bally et al (1987)
with the Bell Telephone 7-m off-set Cassegrain telescope (BTL survey).
The details of the data and analysis method are given in Paper I (Sofue 1995).

\section{2. Dumbbell-Shaped Expanding Molecular Shell}

\sub{2.1. Expanding Ring in Longitude-Velocity Diagrams}

The 200-pc expanding  molecular ring
is recognized in the \lv\ diagrams as the high-velocity
arcs at $\Vlsr \sim \pm 150-200$ \kms, which composes a tilted
ellipse (Bally et al 1978, 1988).
It is known that the vertical extent of this feature is  as large
as $\sim 100$ pc (Sofue 1989).
The vertical extent is significantly greater than that of the nuclear disk
component which contains 85\% of the total emission in the diagram.
Since the ``expanding'' component seen in the \lv\ diagrams
amounts to only 15\% of the total emission in the central $\pm 1\Deg$ region,
it is not a major disk component.
The positive-velocity side of the expanding ring is most clearly visible at
$b\sim 4'$, while the negative-velocity side is most clearly seen
at $b\sim -15'$.
In order to see how the ellipse shows up across its densest parts,
we have combined \lv\ plot at $b=4'$ for positive velocity side and that
at $b=-15'$ for the negative velocity side (Fig. 1).
In the figure the expanding ellipse appears as an almost symmetric
feature around the nucleus.

\ce{-- Fig. 1 --}

\sub{2.2. Latitude-Velocity \bv\ Diagrams}

In order to investigate the vertical structure of these features, we
make use of \bv\ diagrams.
Fig. 2 shows typical \bv\ plots averaged in 10$'$ interval of longitude,
where local components have been subtracted (see Paper I for the analysis).
The expanding ring feature is now
recognized as arched features at $\Vlsr \sim \pm 150 - 180$ \kms.
These arcs are largely extended in the  $b$ direction over
$\sim \pm 20'$ (50 pc),
and are distinguished from the massive disk component which has a
smaller scale height of $\sim \pm 5-10'$.
The arcs are all convex with respect to the center of the diagrams.
The arcs appear to compose an ellipse, which is slightly squeezed near
the galactic plane, representing a dumbbell-shaped expanding
shell feature in the \bv\ space with its neck (equator) at $b=-3'$.
This equator coincides with the plane of the
major molecular disk crossing Sgr A.

\ce{-- Fig. 2 --} 

The velocity width of the shell (velocity difference between the
positive and negative-most edges of the arcs in the \bv\ plot)
appears to vary with longitude.
The width attains maximum at around $l\sim 0^\circ$, and decreases with
the distance from the center.
Although the present data do not cover the regions at
$|l|>1\Deg$, the decrease of the width indicates that the
tangential edges of the expanding
shell are located at  $l=\pm 1\Deg.2~(\pm 180$ pc).

Since  the higher-velocity part of an expanding feature in a position-velocity
plots correspond to a larger-radius part, the \bv\ ellipse (or dumbbell)
features can be most naturally attributed to a cross section of an expanding
spheroidal (or dumbbell-shaped) surface (shell), which we call the
dumbbell-shaped expanding molecular shell (EMS).
{}From the \lv\ diagrams, the EMS is rotating at about 70 \kms\ around the
nucleus with the negative-velocity part being in the near-side.
This rotation speed is much smaller than the galactic rotation velocity,
and may indicate that the gas has been carried from the inner region.

\sub{2.3. Asymmetry about the Galactic Plane,
and Anti-correlation with the Disk Component}

The dumbbell-shape in the \bv\ space has its equator (neck) at $b=-3'$.
which coincides with the plane of the major molecular disk.
However, the \bv\ arcs are not symmetric with respect to the equator:
The lower-$b$ counterpart is missing at $\Vlsr>0$ \kms, while the
upper-$b$ counterpart is missing at $\Vlsr<0$ \kms.
It is remarkable that the latitude of the missing part of the \bv\ arcs
coincide with the latitude where the major disk component is most clearly
seen.
This can be also recognized in \lv\ diagrams:
The expanding feature at negative \vlsr
is missing in the \lv\ diagrams at positive latitudes where the major
molecular arm (Arm I in Paper I) is most prominent.
On the other hand, the feature with positive \vlsr\ is missing at negative
latitudes where Arm II is most prominent.
Fig. 3 shows the relation of the Arms and missing parts of
the expanding molecular features.
In paper I, we interpreted that Arms I and II compose a ring of radius 120 pc
around the Galactic Center, and they are located at opposite
halves of the ring.
The anti-correlation of the expanding shell features with the
Arms can be understood, if the near side (negative velocity side)
of the expanding shell has been interrupted by Arm I which is
in front of the nucleus.
On the other hand, the far side of the shell has been interrupted
by Arm II which is on the far side of the nucleus.

\ce{-- Fig. 3 --} 

\sub{2.4. Parameters for the EMS}

The mass of the whole EMS can be obtained by using a wider-area survey
of Bally et al (1988).
The $^{13}$CO intensity toward $l\sim 1\Deg.2$ is roughly  $\sim 50 $  K \kms.
We use  a recent value of the $^{12}CO$ to H$_2$ conversion factor at
the galactic center,
$0.92\times 10^{20}$ H$_2$ cm$^{-2}$/K \kms (Arimoto et al 1995),
and assuming the $^{12}$CO to $^{13}$CO temperature ratio to be 6.2
(see Paper I).
{}From the ($\Vlsr-l$) diagram we calculated  an  averaged intensity in
$-0\Deg.6 \le b \le +0\Deg.6$,
and estimated the molecular (\htwo\ + He gas)
mass of the EMS as $ M\sim  0.8  \times 10^7 \Msun$.
Since the expansion velocity is about 160 \kms, this leads to a total kinetic
energy of expansion of $ \sim 2 \times10^{54}$ ergs.
Table 1 summarizes the derived parameters for the molecular shell.

\ce{-- Table 1 --}

\sub{2.5. Comparison with Other Data}

A wider view of the expanding-ring feature can be seen in the low-resolution
\lv\ diagram.
Two vertically extended ridges at $l=\pm 1\Deg.2$
are also recognized in the low-resolution integrated intensity maps (Bally et
al 1988).
In Fig. 4  we compare this  global feature with other data as
reproduced from Sofue (1989).
Fig. 4a is a low-resolution \lv\ diagram of the CO line emission.
Fig. 4b shows the CO  intensities integrated within
$-100 \le \Vlsr \le -20$ \kms\  for
$l \le 0\Deg$ and $20 \le \Vlsr \le 100$ \kms\ for $l>0\Deg$, respectively,
as reproduced from Bally et al (1988).
This map represents `cross sections' of the tangential parts of the EMS,
which comprise vertical spurs.
The spur  at $l=1\Deg.2$ is shifted toward the positive latitude side
and can be traced from $b\sim-0\Deg.2$ to $+0\Deg.4$, whereas the spur at
$l=-1\Deg.2$ toward negative latitude, suggesting that the expanding
shell is tilted.

\ce{--  Fig. 4 --} 

Figs. 4c to 4e compare the expanding molecular feature
with the 10 GHz  (Handa et al 1987) and 2.7 GHz (Reich et al 1984)
radio continuum maps as well as the IRAS 60 $\mu$m map (Beichman et al 1985).
Radio continuum maps reveal a number of spurs which emerge from the galactic
plane.
Near the tangential direction of the  EMS at $l \sim 1\Deg.2$,
a spur is visible extending from the galactic plane to
$(l=1\Deg.2,~b=0\Deg.4)$,  which appears to be convex with respect to the
 Galactic center in a consistent sense with the  EMS.
The spurs are also identified on the FIR (far-infrared) emission map of
IRAS at 60 $\mu$m (Beichman et al 1985), indicating that the EMS contains
warm dust.
The observed brightness of the thermal radio emission is  about 100 mK toward
the spur associated with the EMS. If we assume an optically thin
ionized gas of  $T_{\rm e}\sim 10^4$ K and $T_B\sim 0.1$ K at $\nu=10$ GHz,
we obtain an emission measure of $EM \sim 4\times 10^3 ~{\rm pc~cm}^{-6}$.
We then obtain a thermal electron density $n_{\rm e}\sim6~{\rm cm}^{-3}$, and
the total mass of HII gas is derived to be $\sim 4\times 10^5M_\odot$ for the
same geometry as the EMS.
Then the thermal energy involved in the EMS is of the order of  $\sim 10^{51}$
ergs.

A hot plasma of $ \sim10^8$ K has been observed in the Galactic center
in the iron line at 6.7 keV with the Ginga Satellite
(Koyama et al 1989).
The emission region is extended with a size of $1\Deg.8$ (250 pc) about the
galactic center, and  the observed size of
the hot plasma region fits the interior of the EMS.
Considering the angular resolution of the detector and data-sampling grids,
 we may suggest that the  plasma is  surrounded by  the EMS.
However,  the time scale of the expansion of the hot plasma
($\sim$ radius/sound velocity $ \sim 100$ pc/1000 \kms $\sim 10^5$ yr)
is much shorter than the expansion time scale of the molecular feature.
Therefore, the origin of the hot plasma would not be directly related to
the origin of the EMS, but the molecular gas could only act to confine the
plasma in the spheroidal, or possibly cylindrical, space.
Moreover, the thermal energy contained by the hot plasma
($ \sim3 \times 10^{53}$ ergs; Koyama et al 1989; Sofue 1989) is much smaller
than the kinetic energy of the EMS, which implies that the hot plasma
cannot be a driving power for the EMS formation.

\section{3. Discussion}

We have shown that the so-called expanding molecular ring
in the galactic center is a part of a dumbbell-shaped expanding shell
with its equator (neck) in the dense molecular disk.
The shell is squeezed near its equator probably due to interruption by
the dense disk gas near the galactic plane.
The positive latitude and negative-velocity  side (near side of the nucleus)
of the EMS appears to be interrupted by the near-side Arm I, while
the negative-latitude and positive-velocity side (far side) of the EMS
is interrupted by the far-side Arm II.
These structures suggest that the EMS is an expanding molecular
spheroid interacting with the dense molecular disk which comprises
ring and arms.

Fig. 5 illustrates the proposed dumbbell-shaped
spheroid in the $(l, b, \Vlsr)$ space.
This may be transposed to a three-dimensional expanding spheroid
in the $(x, z, y)$ space, where $x$, $z$, and $y$ are axes along the
galactic plane, rotation axis, and depth in the line-of-sight direction,
respectively.
The expanding molecular gas is observed to be distributed in the
intermediate latitude regions of the spheroid.
The gas in the equatorial and polar regions are missing.
The asymmetry of the distribution of gas in the EMS
with respect to the galactic plane
would be somehow related to the fact that the major molecular disk
containing Arms I and II is tilted by 5\deg\ from the galactic plane.

\ce{-- Fig. 5 --}

The total kinetic energy of the shell is approximately $2\times 10^{54}$
ergs, which is equivalent to energy released by
$\sim 10^3$ type II SN explosions.
It is likely that the Galactic center experienced a phase of intense
star formation in the past $\sim 10^6$ years.
Suppose that a coherent star formation containing some $10^3$ SN explosions
(mini burst) lasted for a period short enough compared to the EMS's life
time, e.g., within $\sim 10^5$ yr at the central part of the nuclear disk
some $10^6$ yr ago.
Then, subsequent SN explosions would have formed many shocked shells.
Soon after the explosions, the shock fronts would have accumulated on
a single large shell expanding into the disk and halo.
Such a shocked shell would become deformed by the disk-to-halo density
gradient, and would easily attain a dumbbell shape with the neck
in the disk plane.

Simulations of a shock front evolution in a disk-halo
system have shown that an initially spherical shocked shell becomes
elongated in the direction of the halo and then attain a dumbbell shape
(Sofue 1984, 1994).
If the gravity, which is stronger in the direction
perpendicular to the galactic plane, as well as the fact that the disk gas
before repulsion  was rotating with the centrifugal force,
are taken into account, the oblateness  of the EMS could be accounted for.

Finally, we comment on another model of the expanding molecular feature.
The closed orbit model based on the ``parallelogram" in the
\lv\ plot, as proposed by Binney et al (1991),
would be an alternative promising possibility to explain the present
expanding molecular features in the \lv\ plot.
If the model  is successful in reproducing the rigid rotation features of
the disk component, which
shares 85\% mass of the total emission in the \lv\ plot (Paper I) of the
presently discussed area, this  model would be attractive, since most of
the \lv\ features can be understood under a unified theoretical
scheme of gas dynamics in a central bar potential.
It is also an open question about this model, if it can
explain the three dimensional structure as pointed out in this paper.

\vv
{\bf Acknowledgement}: The author would like to express his sincere thanks
to Dr. John Bally for making him available with the molecular line data
in a machine-readable format.

\section{References}

\parskip=0pt
\def\r{\hangindent=1pc \hangafter=1 \no}

\r Arimoto, N., Sofue, Y., Tsujimoto, T. 1995, submitted to AA.

\r{Bally, J., Stark, A.A., Wilson, R.W., and Henkel, C. 1987, ApJ Suppl 65,
13.}

\r{Bally, J., Stark, A.A., Wilson, R.W., and Henkel, C. 1988, ApJ 324, 223.}

\r{Beichman, C.A., Neugebauer, G., Habing, H.J., Clegg, P.E., and
Chester, T.J. (ed.) 1985, IRAS Explanatory Supplement, JPL D-1855
(Jet Propulsion Laboratory, Pasadena).}

\r Binney, J.J., Gerhard, O.E., Stark, A.A., Bally, J., Uchida, K.I., 1991
MNRAS 252, 210.

\r{Handa, T., Sofue, Y., Nakai, N. Inoue, M., and Hirabayashi, H. 1987,
PASJ 39, 709.}

\r{Kaifu, N., Iguchi, T., and Kato, T. 1974, PASJ 26, 117.}

\r{Kaifu, N., Kato, T., and Iguchi, T. 1972, Nature 238, 105.}

\r{Koyama, K., Awaki, H., Kunieda, H., Takano,S., Tawara, S., Yamanuchi, S.,
Hatsukade, I., and Nagase, F. 1989, Nature  339, 603.}

\r{Reich, W., F{\"u}rst, E., Steffen, P., Reif, K., and Haslam, C.G.T. 1984,
AA Suppl 58, 197.}

\r{Scoville, N.Z. 1972, ApJ  175, L127.}

\r{Sofue, Y. 1984, PASJ 36, 539.}

\r Sofue, Y.  1989,  Ap. Let.  Com. 28, 1 

\r Sofue, Y.  1994, ApJ.L.  431, L91.

\r Sofue, Y.  1995, PASJ, in press (Paper I).

\r{Sofue, Y., and Handa, T. 1984, Nature 310, 568.}

\endpage

\settabs 2 \columns

\no Table 1: Dumbbell-shaped expanding  molecular shell$^*$.
\v
\hrule
\v
\+ Radius\dotfill      & $1\Deg.2$=180  ($\pm 5$) pc      \cr
\+ Vertical extent\dotfill & $\sim \pm 50$ pc \cr
\+ Thickness of the shell \dotfill   & $\sim15$ pc       \cr
\+ Rotation velocity\dotfill  & 70  ($\pm 10$) \kms         \cr
\+ Expansion velocity\dotfill & 160  ($\pm 5$) \kms        \cr
\+ Molecular gas mass\dotfill & $ \sim 0.8\times 10^7 \Msun $  \cr
\+ Kinetic energy\dotfill & $ \sim 2 \times10^{54}$ ergs \cr
\v

\hrule
\no $*$ The distance to the galactic center is taken to be 8.5 kpc.

\no $\dagger$ 10-keV ($1.2\times10^8$ K) thin hot plasma detected by
Koyama et al (1989).

\endpage
\section{Figure Captions}
\vv

Fig. 1: A composite \lv\ diagram by combining data at $b=+4'$ for the upper
half ($\Vlsr>0$ \kms) and that at $b=-15'$ for the lower half ($\Vlsr<0$ \kms).
The map has been smoothed by a Gaussian function with a half-width
$1' \times 2.7$ \kms.
This diagram shows the expanding ring feature at its clearest part.
Contours are in unit of K \ta\ at levels
$0.1\times$(1, 2, 3, 4, 5, 6, 8, 12, 16, 20, 25).
The local and foreground CO emissions have been subtracted by applying
the ``pressing method'' to the  BTL cube data (Bally et al 1987)
(see Paper I for the procedure).

\vv
Fig. 2:
(a) Typical \bv\ diagrams after subtraction of local/foreground emission.
The expanding molecular shell can be fitted by a dumbbell-shape
as illustrated by the full lines with its neck (equator) at $b=-3'$.
Contours are in unit of K \ta\ at levels
$0.1\times$(1, 2, 3, ..., 9, 10, 12, 14, 16, 18, 20, 25, 30, 35, 40).

(b) All \bv\ diagrams at every 5$'$ longitude interval in grey scale.
Each diagram has the same $b$ and \vlsr\ coverage as in (a).
The longitude decreases from the left-bottom at $l=+1\Deg$
to the right, toward top, and ends at the right-top panel at $l=-1\Deg$.

\vv
Fig. 3: \lv\ diagrams at $b=+2$ to $+4'$ and at $b=-10$ to $-8'$.
This diagrams show an anti-correlation of the disk components Arms I and
II with the missing parts of the expanding features at positive and
negative latitude sides, respectively.

\vv
Fig. 4: Comparison in the same scale of $^{13}$CO$(J=1-0)$ data with
radio and FIR maps (reproduced from Sofue 1989):

(a) The \lv\ diagram of the $^{13}$CO gas (reproduced from Bally et al 1988).

(b) CO intensity distribution integrated in the velocity ranges from 20 to
100 \kms\ for $l \ge 0\Deg$, and from $-20$ to $-200$ \kms\ for $ l \le 0\Deg$
(Bally et al 1988).

(c) A radio continuum map at 2.7 GHz (Reich et al
1984). The background emission of scale sizes greater than $ 0\Deg.4$
has been subtracted,  and the map is smoothed to $6'$ resolution.

(d) The same as (c) but at 10 GHz (Handa et al 1987).

(e) The same as (c) but at 60 $\mu$m made from the IRAS data (Beichmann
1985).

\vv
Fig. 5: Schematic view of the dumbbell-shaped shell
in the $(l,b,\Vlsr)$ space, which may be transposed to a similar
shaped three-dimensional expanding spheroid, which is rotating at
70 \kms and expanding at 160 \kms.
The molecular gas is distributed in the intermediate-latitude regions
of the spheroid, and is missing in the equatorial and polar regions.

\bye